\documentstyle[aps,multicol,epsf]{revtex}
\newcommand{\beq}{\begin{equation}}
\newcommand{\eeq}{\end{equation}}
\newcommand{\pa}{\partial}
\newcommand{\al}{\alpha}

\begin{document}
\draft

\title{Possible Stratification Mechanism in Granular Mixtures}

\author{Hern\'an A. Makse, Pierre Cizeau,
and H. Eugene Stanley}

\address{Center for Polymer Studies and Physics Dept., Boston
University, Boston, MA 02215 USA}

\date{Phys. Rev. Lett. {\bf 78}, 3298-3301 (1997)}

\maketitle
\begin{abstract} 
We propose a mechanism to explain what occurs when a mixture of grains
of different sizes and different shapes (i.e. different repose angles )
is poured into a quasi-two-dimensional cell. Specifically, we develop a
model that displays spontaneous stratification of the large and small grains in
alternating layers. We find that the key requirement for
stratification is a difference in the repose angles of the two pure
species, a prediction confirmed by experimental findings. We also
identify a kink mechanism that appears to describe essential aspects of
the dynamics of stratification.
\end{abstract}

\pacs{PACS Numbers: 05.40+j, 46.10+z, 64.75+g}
\begin{multicols}{2}

\narrowtext

Granular materials 
exhibit many unusual properties \cite{review}, 
such as size segregation, when
exposed to external vibrations or rotations \cite{brazilnut}.  
Size segregation is also observed when a mixture of grains of different
size
is poured onto a pile \cite{segregation},
the large grains are found preferentially near the bottom of the pile,
while the small grains are found near the top.
Recently,
it was found \cite{makse}
that 
when a mixture of grains of
different sizes and shapes is poured between two vertical slabs
separated by $\approx 5$~mm, there appears a spontaneous 
stratification, with alternating layers of small and large
grains parallel to the surface of the sandpile. Additionally, there is
an overall tendency for the large and small grains to 
spontaneous segregate in different regions of the cell
\cite{segregation,makse}. 

Very recently, Boutreux and de Gennes (BdG) \cite{degennes} treated
theoretically the case of granular flow made of two species. They based
their work on a set of coupled convection equations to govern the flow
of rolling grains and their interaction with the sandpile, introduced
earlier by Bouchaud {\it et al.} in the case of a single-species
sandpile \cite{bouchaud}. BdG reproduced the phenomenon of
segregation, but an understanding of stratification is
lacking.

Here we seek to understand segregation and stratification in
the conditions of \cite{makse}, where the two species have different
size and different shape. We first introduce a discrete model to give a
clear picture of the phenomenology, and then develop a continuum
approach. In agreement with the experimental findings \cite{makse}, we
find that segregation is related to the difference of size of the
grains, and stratification, to the difference in repose angles of
the two pure species.

In the discrete model, the sandpile is built on a lattice, where the
grains have the same horizontal size as the lattice spacing and two
different heights, $H_1$ and $H_2 > H_1$. Each grain belongs to one of
two phases: a {\it static phase\/} (if the grain is part of the solid
sandpile) and a {\it rolling phase\/} (if the grain is not part of the
sandpile but rolls downward with a constant drift velocity)
\cite{degennes,bouchaud,degennes-french}.  The local slope $s_i \equiv
h_{i} - h_{i+1}$ of the static grains is the variable controlling the
dynamics of the rolling grains, where $h_i$ is the height of the
sandpile at column $i$.

At each time step, we deposit at the top of column $1$ of the pile $N_1$
small grains plus $N_2$ large grains; these grains belong to the {\it
rolling phase}.  One rolling grain per column of each species interacts
with the surface at each time step, and can be converted from the
rolling phase to the {\it static phase}. The remaining rolling grains
are convected downward with unit drift velocity---i.e., they all move
to the next column at each time step.

\begin{figure}[tbp]
\centerline{ (a)\vbox{  \hbox{\epsfxsize=4.2cm
 \epsfbox{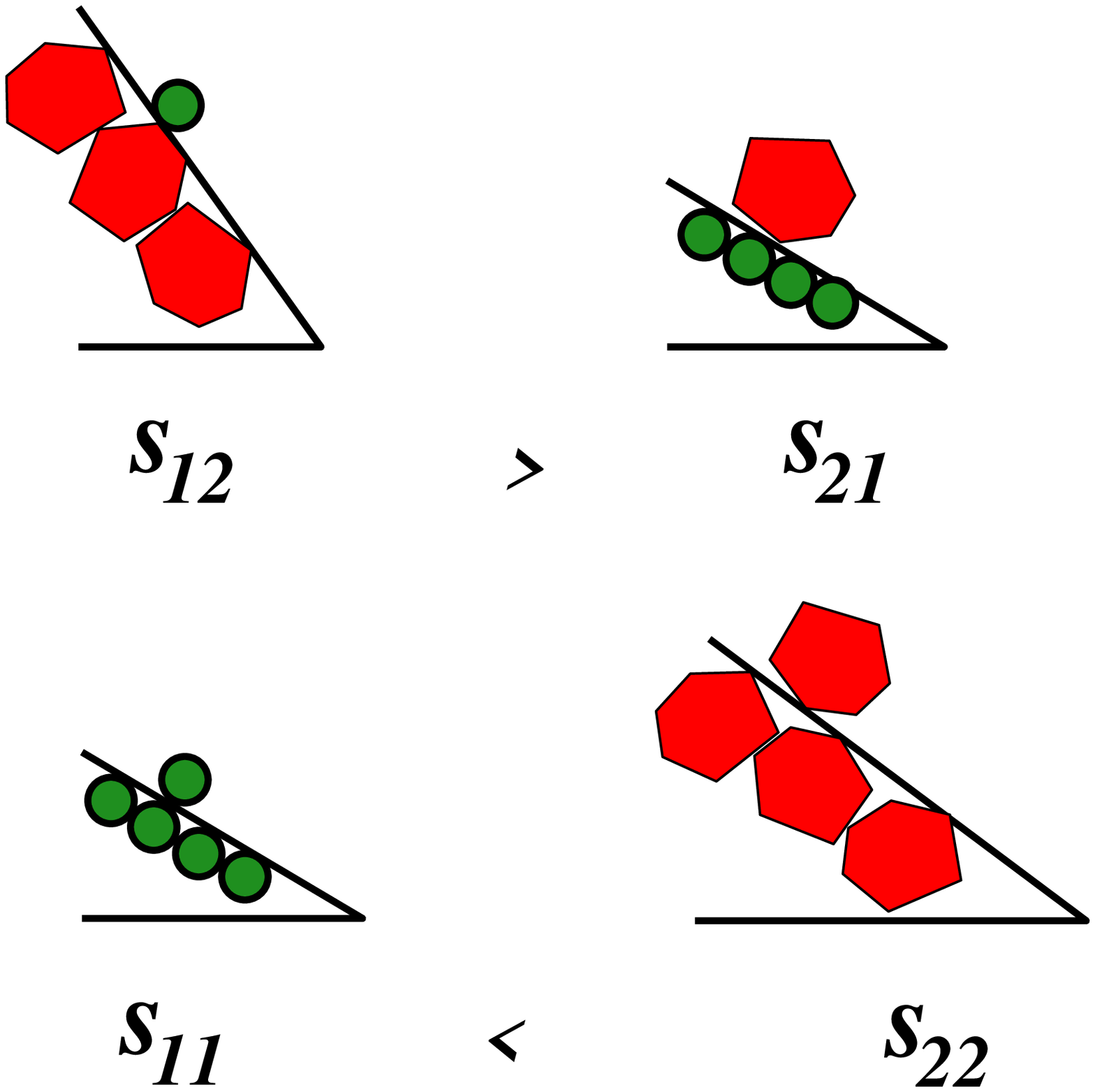} }}
\vspace{1cm}}
\centerline{
(b) \vbox{   \hbox{\epsfxsize=5cm \epsfbox{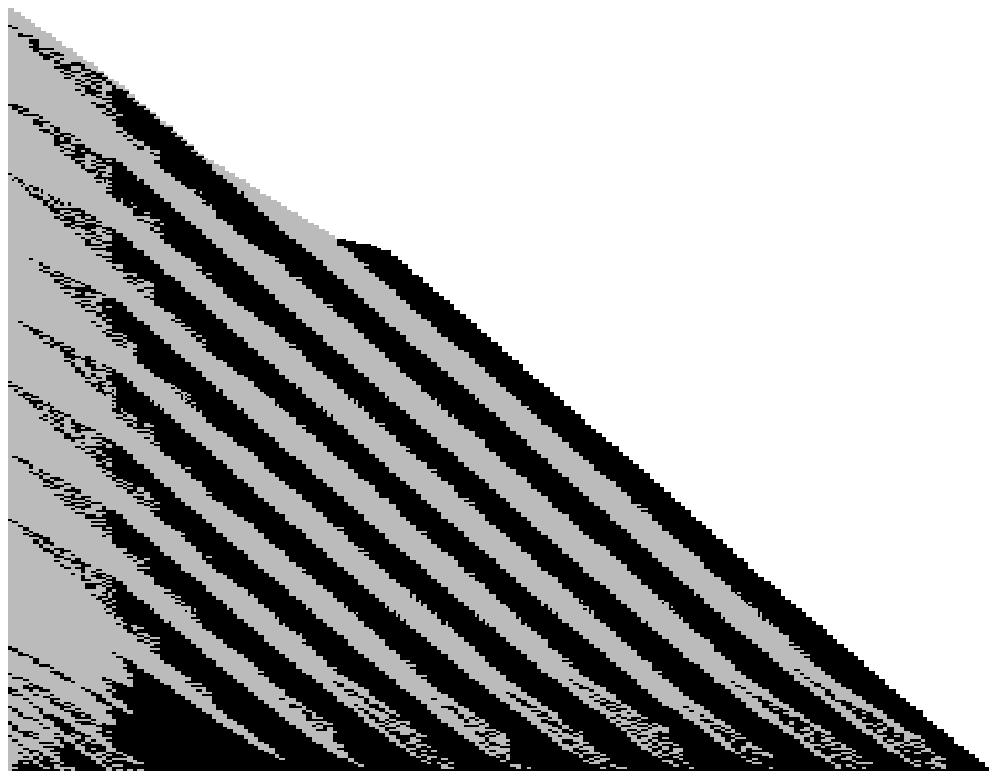}}
     }
       \vspace{1cm}    }
\narrowtext
\caption{(a) The slopes of repose $s_{\alpha \beta}$ depend on
the composition of grains at the surface of the pile, and are chosen
according of the four possible interactions between small and large
grains. The slopes of repose satisfy $s_{21} < s_{11} < s_{22}< s_{12}$
(see discussion in text).  (b) Result obtained with the discrete model
(for $H_1=1$, $H_2=2$, $s_{21}=2$, $s_{11}=6$, $s_{22}=7$, $s_{12}=10$,
$N_1=20$, and $N_2=10$).}
\label{discrete}
\end{figure}

The dynamics of each rolling grain interacting with the sandpile surface
is governed by its repose angle (the maximum angle below which a rolling
grain is converted into a static grain) \cite{bouchaud,bagnold}.
We note that the repose angle depends on the local composition on the
surface, so we define $\theta_{\alpha\beta}$ as the repose angle of a
rolling grain of type $\alpha$ on a surface with local grains of type
$\beta$. We choose $\theta_{21}<\theta_{12}$ to take into account that
large grains roll more easily on top of small grains than small grains
roll on top of large grains (since the surface ``looks'' smoother for
large grains rolling on top of small grains, see Fig.  \ref{discrete}a).
The repose angles of pure species
$\theta_{\alpha\alpha}$ lie between $\theta_{21}$ and $\theta_{12}$.

The stratification experiments \cite{makse} use a mixture of grains
of different shapes (smaller ``less faceted'' grains and larger ``more
faceted'' grains). The repose angle of the smaller pure species is then
smaller than the repose angle of the large pure species---i.e.,
$\theta_{11} < \theta_{22}$.  To mimic the experimental conditions for
stratification \cite{makse}, we set $\theta_{21} < \theta_{11} <
\theta_{22}<\theta_{12}$.  We notice that the condition $\theta_{21} <
\theta_{12}$ is a consequence of the different size of the species,
while the condition $\theta_{11} \neq \theta_{22}$ is achieved, e.g.,
for mixtures of grains with different shapes.

At each time step, the rolling grain interacting with the sandpile
surface at coordinate $i$ will either stop---by being converted into a
static grain---if the local slope of the surface $h_{i} - h_{i+1} <
s_{\alpha\beta}\equiv \tan \theta_{\alpha\beta}$, or will continue to
roll (together with the remaining rolling grains) to column $i+1$ if
$h_{i} - h_{i+1} \ge s_{\alpha \beta}$. We iterate this algorithm to form
a large sandpile of typically $10^5$ grains.

Figure \ref{discrete}b shows the resulting morphology.  The
stratification is qualitatively the same as that found experimentally
\cite{makse}, not only in regard to the {\it statics} of the sandpile
(seen in Fig. \ref{discrete}b), but also in regard to the {\it
dynamics}.  After a pair of static layers is formed with a layer of
large grains on top of a layer of small grains, the angle of the
sandpile is close to $\theta_{22}$.  Since the surface of the sandpile
is made of large grains and $\theta_{22}<\theta_{12}$, a thin layer of
small grains is trapped on top of the layer of large grains.  These
small grains smooth the surface without changing significantly the
sandpile angle, and allow rolling small grains to go further down (since
$\theta_{11}<\theta_{22}$). The large grains are rolling down on this
thin layer of small grains without being captured
($\theta_{21}<\theta_{22}$), and are the first to reach the bottom of
the sandpile, giving rise to segregation.  When the flow reaches
the base of the pile, the grains develop a profile which displays a
``kink'' where the grains are stopped: the small grains are stopped
first since $\theta_{21} < \theta_{12}$, and a pair of layers begins to
form, with the small grains underneath the large grains.  The kink moves
upward (in the opposite direction to the flow of grains) until it
reaches the top of the pile and a complete pair of layers has been
formed.

If, on the other hand, we consider 
$\theta_{22} < \theta_{11}$ in the model
(corresponding to a
mixture of smaller ``more faceted'' grains, and larger ``less faceted''
grains) we find segregation but no stratification. Thus, the
control parameter for stratification appears to be the difference
in the repose angle of the pure species.

As seen in Fig.~\ref{discrete}b, before the layers appear there is an
initial regime where only segregation is found.  At the onset of
the instability leading to stratification, a few large grains are
captured on top of the region of small grains near the center of the
pile where the angle of the pile is $\theta \simeq \theta_{11}$.  The
repose angle for large grains is now $\theta_{22}$.  Thus, if $\theta
\simeq \theta_{11} < \theta_{22}$, more large grains can be trapped
(since the angle of the surface is smaller than the repose angle),
leading to the first sublayer of large grains and then to
stratification.  On the other hand, if $\theta \simeq \theta_{11} >
\theta_{22}$, no more large grains are captured, the fluctuation
disappears, and the segregation profile remains stable.  Thus this
picture suggests that, in agreement with \cite{makse}, the
segregation profile observed in the initial regime is ``stable'' so
long as $\theta_{22}< \theta_{11}$, and unstable (evolving to
stratification for large enough systems) when $\theta_{11} <
\theta_{22}$.

To offer further insight into the physical mechanism of
stratification, we now develop a continuum approach in the spirit
of Refs.~\cite{degennes,bouchaud,degennes-french}.  The variables are
the two thicknesses of rolling grains $R_\al(x,t)$, with $\al =1,2$
respectively for small and large grains, the height of the sandpile
$h(x,t)$, and the volume fraction of grains of $\beta$ type in the
static phase $\phi_\beta(x,t)$.
The equations of motion are  \cite{degennes,BdG}
\begin{mathletters}
\label{eq:R_et_h}
\begin{eqnarray}
\frac{\pa R_\al}{\pa t} & = & -v_\al \frac{\pa
R_\al}{\pa x} + \Gamma_\alpha \\
\frac{\pa h}{\pa t}  & = & -\sum_\alpha \Gamma_\alpha
\end{eqnarray} 
\end{mathletters}
Here $v_\al$ is the downhill convection velocity of species $\al$, and
$\Gamma_\alpha$ describes the interaction of the rolling grains with the
surface---i.e., how rolling grains are stopped and become part of the
sandpile (capture), and how grains of the sandpile can enter the rolling
phase (amplification).  The concentrations are given by
$\phi_1+\phi_2=1$, and $\phi_\alpha=-\Gamma_\alpha/(\pa h /\pa t)$.  

\begin{figure}[htbp]
\centerline { \vbox{ \hbox
{\epsfxsize=6.5cm \epsfbox{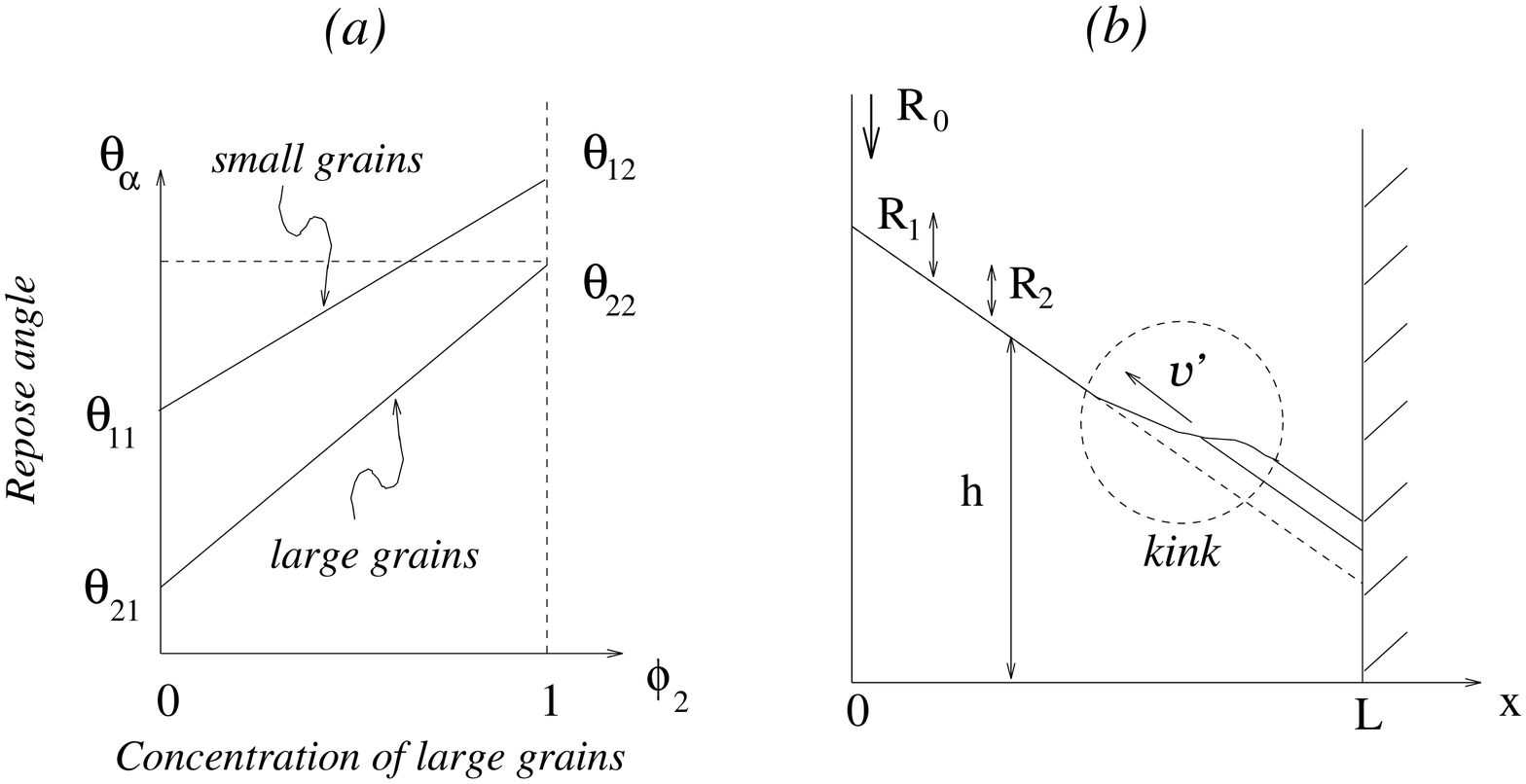} }}}
\centerline { \vbox{ \hbox
{\epsfxsize=5.8cm \epsfysize=5.6cm \epsfbox{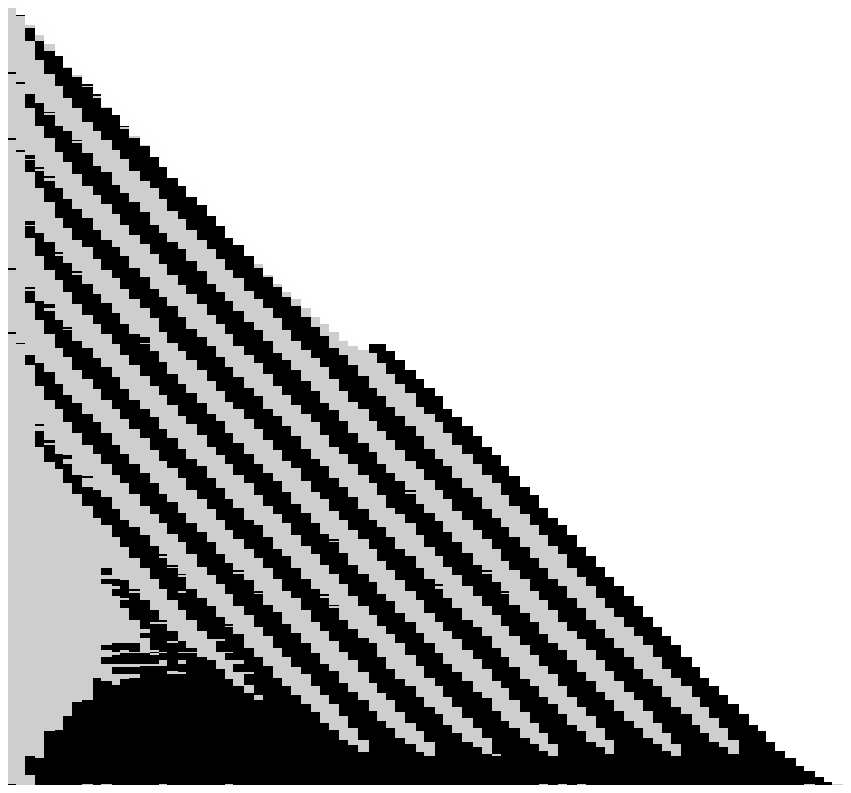} }}}
\narrowtext
\vspace{0.5cm}
\caption{(a) Dependence of the repose angle for the two types of rolling 
grains on the concentration of the surface of large grains $\phi_2$. An
essential ingredient to obtain stratification is that $\theta_{22} >
\theta_{11}$. For the numerical integration, we use the linear
interpolation between $\phi_2=0$ and $\phi_2=1$ as plotted here.  (b)
Picture with the different quantities appearing in the text. The
dash-circled zone is the ``kink.''  (c) Resulting morphology of the
numerical integration of the continuum equations. The parameters used
are $\tan\theta_{11}=1$, $\tan\theta_{22}=1.1$, $\tan\theta_{12}=1.4$,
$\tan\theta_{21}=0.7$~, $\gamma_{1}= \gamma_{2}=0.8$, and
$v_1=v_2=1$.} 
\label{conti}
\end{figure}

As in the discrete model, we focus on the dependence of the repose angle
on the composition of the surface $\phi_\beta(x,t)$.  The repose angle
$\theta_\alpha$ of each type of rolling grain is now a continuous
function of the composition of the surface $\theta_\alpha =
\theta_\alpha(\phi_\beta)$ (see Fig.~\ref{conti}a). The repose angle
$\theta_{\al\beta}$ defined for the discrete model is now
$\theta_\al(\phi_\beta)$ with $\phi_\beta=1$.  We propose that
$\Gamma_\alpha = \Gamma_\alpha (R_\alpha, \phi_\beta,
\theta_{\mbox{\scriptsize loc}})$ obeys the relation
\beq
\label{eq:M}
\Gamma_\alpha\equiv \left \{ 
\begin{array}{ll}
\gamma_\alpha 
(\theta_{\mbox{\scriptsize loc}} -\theta_\al(\phi_\beta))
 ~R_\alpha& ~~~~~ \mbox{ if $\theta_{\mbox{\scriptsize loc}} < 
\theta_\al(\phi_\beta) $} \\
\gamma_\alpha ~ \phi_\alpha
(\theta_{\mbox{\scriptsize loc}} -\theta_\al(\phi_\beta))
 ~R_\alpha& ~~~~~ \mbox{ if $\theta_{\mbox{\scriptsize loc}} > 
\theta_\al(\phi_\beta) $} 
\end{array}
\right .
\eeq 
where 
$\theta_{\mbox{\scriptsize loc}} (x,t) \equiv -\pa h/\pa x$ is the
local surface angle.  The parameter $\gamma_\alpha$ represents the
effectiveness of the interaction: $v_\alpha/\gamma_\alpha \sim d_\alpha$
(where $d_\alpha$ is the linear size of the grain) is the length scale
on which a rolling grain will interact significantly with the surface
when $\theta_{\mbox{\scriptsize loc}}$ is slightly
different from the repose angle
\cite{degennes-french}.

The interaction term $\Gamma_\alpha$ includes two types of process
\cite{degennes}.  {\bf \it(a) Capture}: as in the discrete model,
rolling grains are captured if $\theta_{\mbox{\scriptsize loc}} <
\theta_\al(\phi_\beta)$.
The capture is proportional to the number of grains interacting with the
surface, and so is proportional to $R_\al$. {\bf \it(b) Amplification}:
if $\theta_{\mbox{\scriptsize loc}}>\theta_\al(\phi_\beta)$, then some
static grains of the sandpile are converted into rolling grains. This
conversion of $\al$ type grains is proportional to their surface
concentration $\phi_\al$, and to the number $R_\al$ of rolling grains
acting in amplification.

We next solve Eqs.~(\ref{eq:R_et_h})--(\ref{eq:M}) numerically
\cite{conditions}.  The results, shown in Fig.~\ref{conti}c, are
qualitatively similar to the discrete model \cite{conditions}. We find
stratification whenever the repose angle has the qualitative
behavior shown in Fig.  \ref{conti}a, the key requirement being
$\theta_{22}>\theta_{11}$.
We also find a ``kink,'' corresponding to the growth of the new pair of
static layers, with a well-defined steady-state profile and upward
velocity.

To find the conditions under which stratification occurs, we first
calculate the steady-state solution of the full set of Eqs.
(\ref{eq:R_et_h})-(\ref{eq:M}), and then study its stability under
perturbations. To describe the experimental situation, we consider a 2-D
cell with vertical walls at $x=0$ and $x=L$ \cite{degennes}.  We assume
that the difference $\psi \equiv \theta_1(\phi_2) - \theta_2(\phi_2) $
is independent of the concentration $\phi_2$, and we set $v_1=v_2\equiv
v$, and $\gamma_1=\gamma_2\equiv\gamma$ (Fig. \ref{conti}b).  We seek a
steady-state solution, where the profiles of the sandpile and of the
rolling grains are conserved in time. Thus, stratification cannot be
observed for this solution. The conservation of the grains gives
$\partial \overline{h} / \partial t = vR^0 / L$, and we impose $\partial
\overline{R}_\alpha / \partial t = 0$, with boundary conditions
$\overline{R}_\alpha (0) = R_\alpha^0 = R^0/2$ and $\overline{R}_\alpha
(L) =0$.

The steady-state solution of (\ref{eq:R_et_h})-(\ref{eq:M}) shows almost
total segregation.  At the {\it upper} part of the pile, for
$0<x<x_m$, with $x_m \equiv L/2-v/(\gamma \psi)$, only small grains are
present ($\overline{\phi}_1(x)=1$, and $\overline{\phi}_2(x) = 0$), and
the profiles are
\begin{mathletters}
\label{steady-up}
\beq
\overline{R}_1(x) = R^0\left(\frac{1}{2} -\frac{x}{L}\right), \>\>\>\> 
\overline{R}_2(x) = \frac{R^0}{2} 
\eeq
\beq
\overline{\theta}_{\mbox{\scriptsize loc}}(x)-\theta_{11} 
=\frac{-v/\gamma}{ L/2-x}.
\eeq
\end{mathletters}

At the {\it lower} part of the pile ($x_m < x < L$), we find that, after
a small region of size of the order of $v/(\gamma \psi)$, mainly
large grains are present, and the profiles are 
\begin{mathletters}
\label{steady-down}
\begin{eqnarray}
\overline{\phi}_1(x) & = & \exp\left[{-\frac{\gamma\psi}{v}(x-x_m)}\right] \\
\overline{R}_1(x)& =& \frac{2v}{\gamma \psi L} \overline{\phi}_1(x)
\overline R (x) \\
\overline \theta_{\mbox{\scriptsize loc}} (x)-\theta_{22} &=& -m \overline{\phi}_1(x)-\frac{v}{\gamma
(L-x)}.
\end{eqnarray}
\end{mathletters}
Here $\overline{R}\equiv \overline{R}_1+\overline{R}_2=R^0(1-x/L)$, and
$m\equiv \theta_{22}-\theta_{21}=\theta_{12}-\theta_{11}$.


To analyze the stability of the steady-state solution (\ref{steady-up})
and (\ref{steady-down}) for the different phenomenologic parameters, we
impose the steady-state solution as the initial condition, and then we
look numerically for the stability of the profile under
perturbations. For $\theta_{11}>\theta_{22}$, the steady state solution
is {\it stable\/}: in this case, only segregation is observed, and
the sandpile conserves in time the profiles (\ref{steady-up}) and
(\ref{steady-down}). For $\theta_{11}<\theta_{22}$, the steady-state
solution is {\it unstable\/} (evolving to stratification), just as
in \cite{makse}.

To gain insight about the ``kink'' mechanism, we look for a possible
steady-state solution for the shape of the kink assuming that (i) far
below and above the kink, the sandpile has a constant angle $\theta_0$;
(ii) the lowest part of the kink is made only of small grains, so that
large rolling grains are not captured, and the top part is made only of
large grains.

To suppose the existence of a stationary solution for the kink implies
that $R_1(x,t)$ and $f(x,t)\equiv h(x,t)+\theta_0 x$ are functions only
of $u\equiv x+v_\uparrow t$, where $v_\uparrow$ is the uphill speed of
the kink. For the lowest part of the kink, as only small grains are
captured ($\phi_1(u)=1$, $R_2(u)=R^0/2$), Eqs.  (\ref{eq:R_et_h}) reduce
to equations for $R_1(u)$ and $f(u)$. We obtain the shape of the low
part of the kink: for $u \le 0$, $f(u)=0$ and for $u>0$, $f(u)$ obeys
\beq
\label{lower_kink}
-\frac{1}{w}\log\left(1-\frac{2wf}{R^0}\right) =
\frac{\gamma}{v_\uparrow}(f-\delta_1 u), 
\eeq 
where $\delta_1 \equiv \theta_0-\theta_{11}$, and $w \equiv
v_\uparrow/(v+v_\uparrow)$.  Then the lower layer of the kink is
characterized by a linear dependence $f(u) \propto u \delta_1$ for $u\ll
R^0 / (2 w \delta_1)$, plus logarithmic corrections near the boundary
with the upper layer of large grains.  This solution is no longer valid
when the angle of the surface reaches $\theta_{21}$, and the large
grains start to be captured.  We note that this stationary solution
exists only when $\delta_1 >0$.

The solution of the equations for the highest part of the kink where
only large grains are present can be obtained in the same way and is
\beq
\label{upper_kink}
f(u) = \left({R^0\over w}\right) \left[1-\exp \left({w\gamma\delta_2
      u\over v_\uparrow}\right) \right],
\eeq
where $\delta_2\equiv\theta_0-\theta_{22}$. We then find that the shape
of the upper part of the kink is exponential, and exists only for
$\delta_2<0$.

Thus we see that the existence of the stationary solution for the kink
implies that $\theta_{11}<\theta_0<\theta_{22}$: the sandpile is built
on an angle intermediate between the two repose angles of the pure
species, and the repose angle of the small grains must be smaller than
the repose angle of the large grains --- in agreement with experiments
\cite{makse}, and the stability analysis performed above.

The layer thickness $\lambda$ is $R^0/w$ (see Eq.  \ref{upper_kink}),
which is a consequence of the conservation law stating that all the
rolling grains are stopped at the kink \cite{makse}.  Furthermore, Eq.
(\ref{lower_kink}) implies $\gamma w f/v_\uparrow \sim 1$. For $f \sim R^0/w$,
this gives $v_\uparrow \propto \gamma R^0$, so that, for $R^0\neq 0$, we obtain
\beq
\label{eq:lambda}
\lambda ~\propto ~c~ v/\gamma ~+~ R^0.
\eeq
where $c$ is a numerical constant that does not depend on $v$, $\gamma$,
or $R^0$. This relation, which we verify numerically, is relevant since
$v/\gamma$ and $R^0$ are both of the order of the diameter of the
grains.

The typical size of the initial regime of segregation, $L_x$, observed
prior to stratification when $\delta\equiv\theta_{22} - \theta_{11}>0$
(Fig.~\ref{conti}c), can be calculated as follows.  The condition for
the appearance of a first layer of large grains on top of the region of
small grains near the center of the pile is that capture of large grains
must be larger than capture of small grains, i.e. $|\Gamma_2| >
|\Gamma_1|$, where $|\Gamma_1| = \gamma m R_1$, and $|\Gamma_2| = \gamma
\delta R_2$.  Assuming that the solution (\ref{steady-up}) is valid for
the initial segregation regime, we can evaluate $R_1$, and $R_2$ at
$x=x_m$. We obtain
\begin{equation}
\label{e.5}
L_x \simeq \frac{v}{\gamma \psi} \frac{m}{\delta} \frac{R^0}{R^0_2},
\end{equation}
and verify (\ref{e.5}) numerically. 

In conclusion, we develop a mechanism to explain the observed
stratification \cite{makse}. This mechanism is related to the
dependence of the local repose angle on the local surface composition.
We find that stratification occurs only when the repose angle of
the large grains is larger than the repose angle of the small grains
($\theta_{22}>\theta_{11}$, corresponding to large grains rougher than
small grains). The model describes the static picture of the sandpile of
\cite{makse} with alternating layers made of small and large grains, and
also reproduces the dynamics, where the layers are built through a
``kink'' mechanism.  When $\theta_{22}<\theta_{11}$, the model predicts
almost complete segregation, but not stratification.  These results
are in agreement with experiment \cite{makse}.

We thank T. Boutreux, S. Havlin and P. R. King for stimulating
discussions, S. T.  Harrington, R. Sadr, and S. Zapperi for comments on
the manuscript, P.-G. de Gennes for sending us Ref.~\cite{degennes}
prior to publication, and BP for financial support.

\end{multicols} 

\begin{references}

\bibitem{review} H. M. Jaeger and S. R.  Nagel, Science {\bf 255}, 1523
(1992); H. J. Herrmann in {\it Disorder and Granular Media\/},
D. Bideaux and A.  Hansen (eds.) (North-Holland, Amsterdam, 1993).


\bibitem{brazilnut} J. C. Williams, Powder Technol. {\bf 15}, 245 (1976); 
J. A. C. Gallas, H. J. Herrmann, and S. Sokolowski,
Phys. Rev. Lett.  {\bf 69}, 1371 (1992); J. B. Knight {\it et al.},
{\it Ibid.} {\bf 70}, 3728 (1993);  O. Zik {\it et al.}
{\it Ibid.} {\bf 73}, 644 (1994).

\bibitem{segregation}
R. L. Brown, 
{\it J. Inst.  Fuel} {\bf 13}, 15 (1939);
R. A. Bagnold, {\it The physics of blown sand and desert dunes} (Chapman
and Hall, London 1941)


\bibitem{makse}
H. A. Makse, S. Havlin, P. R. King, and H. E. Stanley, Nature  {\bf 386},
379-381 (1997).

\bibitem{degennes} 
T. Boutreux and P.-G. de Gennes, J. Phys. I France {\bf 6}, 1295 
(1996).


\bibitem{bouchaud} J.-P. Bouchaud, M. E. Cates, J. R. Prakash, and
 S. F. Edwards, Phys. Rev. Lett. {\bf 74}, 1982 (1995);
 J. Phys. I France {\bf 4}, 1383 (1994).

\bibitem{degennes-french} P.-G. de Gennes, C. R. Acad. Sci. (Paris) {\bf
321~[IIb]}, 501 (1995).




\bibitem{bagnold} R. A. Bagnold, Proc. Roy. Soc. London {\bf A 295}, 219
(1966).




\bibitem{BdG} BdG \protect \cite{degennes} proposed
Eqs. (\protect\ref{eq:R_et_h}).  They did not consider the dependence of
the repose angle on the surface composition, and used a different
interaction term $\Gamma_\alpha$, taking into account another type of
amplification process (where static grains of one type are amplified by
rolling grains of the other type).  We postpone treating this
amplification for a subsequent work.

\bibitem{conditions} In the conditions of the experiment \cite{makse},
where an equal volume of the two species is poured at the left side of
the cell, the boundary conditions are $ R_\al(0,t)=R^0/2$.  Equations
(\protect \ref{eq:R_et_h})-(\protect \ref{eq:M}) are meaningful if all
the rolling grains interact all the time with the surface.  This implies
that we are in the limit of low flow, i.e. $R_\al \simeq d_\al$.


\end{references}
\end{document}